# Magnetization Transfer-Mediated MR Fingerprinting

**Authors**: Daniel J. West[1]*, Gastao Cruz[1], Rui P.A.G. Teixeira[1,2], Torben Schneider[3], Jacques-Donald Tournier[1,2], Joseph V. Hajnal[1,2], Claudia Prieto[1], Shaihan J. Malik[1,2]

1. Department of Biomedical Engineering, School of Biomedical Engineering and Imaging Sciences, King's College London, St. Thomas' Hospital, London, SE1 7EH, United Kingdom
2. Centre for the Developing Brain, School of Biomedical Engineering and Imaging Sciences, King's College London, St. Thomas' Hospital, London, SE1 7EH, United Kingdom
3. Philips Healthcare, Guildford Business Park Road, Guildford, GU2 8XG, United Kingdom

* Correspondence to Daniel West (daniel.d.west@kcl.ac.uk), Perinatal Imaging and Health, 1st Floor South Wing, St. Thomas' Hospital, London, SE1 7EH, United Kingdom.

**Acknowledgements**: This work was supported by the King's College London & Imperial College London EPSRC Centre for Doctoral Training in Medical Imaging (EP/L015226/1) and the Wellcome/EPSRC Centre for Medical Engineering (WT 203148/Z/16/Z).

**ABSTRACT**

**Purpose**: Magnetization transfer (MT) and inhomogeneous MT (ihMT) contrasts are used in MRI to provide information about macromolecular tissue content. In particular, MT is sensitive to macromolecules and ihMT appears to be specific to myelinated tissue. This study proposes a technique to characterize MT and ihMT properties from a single acquisition, producing both semiquantitative contrast ratios, and quantitative parameter maps.

**Theory and Methods**: Building upon previous work that uses multiband radiofrequency (RF) pulses to efficiently generate ihMT contrast, we propose a cyclic-steady-state approach that cycles between multiband and single-band pulses to boost the achieved contrast. Resultant time-variable signals are reminiscent of a magnetic resonance fingerprinting (MRF) acquisition, except that the signal fluctuations are entirely mediated by magnetization transfer effects. A dictionary-based low-rank inversion method is used to reconstruct the resulting images and to produce both semiquantitative MT ratio (MTR) and ihMT ratio (ihMTR) maps, as well as quantitative parameter estimates corresponding to an ihMT tissue model.

**Results**: Phantom and *in vivo* brain data acquired at 1.5T demonstrate the expected contrast trends, with ihMTR maps showing contrast more specific to white matter (WM), as has been reported by others. Quantitative estimation of semisolid fraction and dipolar $T_1$ was also possible and yielded measurements consistent with literature values in the brain.

**Conclusions**: By cycling between multiband and single-band pulses, an entirely magnetization transfer mediated 'fingerprinting' method was demonstrated. This proof-of-concept approach can be used to generate semiquantitative maps and quantitatively estimate some macromolecular specific tissue parameters.

# 1. INTRODUCTION

Magnetization transfer (MT) and inhomogeneous MT (ihMT) contrasts are used in MRI to provide information about macromolecular tissue content. In particular, MT is sensitive to macromolecules and ihMT appears to be specific to substances with non-zero dipolar magnetization order, including the lipid bilayers that form myelin (1–4). Recently, ihMT measurements have: compared favorably against other myelin imaging metrics (5,6); correlated strongly with fluorescence microscopy findings (7); and provided sensitivity for the assessment of demyelinating conditions, such as multiple sclerosis (8). MT and ihMT contrasts are usually obtained by sequences with off-resonance radiofrequency (RF) saturation pulses (that only affect the semisolid pool), followed by readout periods for measurement. For example, ihMT gradient echo methods have been developed and optimized to achieve high-resolution whole-brain imaging at 1.5T (9).

The ihMT effect arises because semisolid magnetization can be modelled as containing pools of both Zeeman and dipolar order that can be made to exchange by the presence of off-resonant RF irradiation; dual frequency saturation with equal and opposite frequency offsets cancels this interaction. In our previous work, it was shown that ihMT contrast can also be generated using non-selective multiband pulses that perform off-resonance saturation and on-resonance excitation simultaneously (10). These pulses were originally proposed to control for MT effects in variable flip angle techniques by ensuring constant RF power across all flip angles, resulting in more stable relaxometry measurements (11). In order to generate ihMT contrast, images using 2-band excitation (one on-resonance and one off-resonance band) leading to Zeeman-dipolar coupling, were compared with 3-band excitation (one on-resonance and two equal and opposite off-resonance bands) in which this coupling is cancelled. Use of the same total power results in the same classical MT effect, for tissues where dipolar order can be neglected.

Resultant ihMT ratios (ihMTRs) from this type of sequence are generally small, for example ~4% in white matter (WM), which is similar to other steady-state measurement techniques (9). In order to boost this contrast, Varma *et al*. recently proposed use of a low duty-cycle RF saturation scheme where saturation pulses are concentrated into short time periods with interleaved recovery time during which data are acquired (12). With this as motivation, in this work we propose a modulated cyclic-steady-state sequence employing a balanced steady-state free precession (bSSFP) acquisition with multiband pulses that are alternated over time to create a similar low duty-cycle saturation effect. Each RF pulse contains an on-resonance component, meaning that data is continuously read out, improving the efficiency of encoding.

By holding the on-resonance flip angle constant while periodically alternating off-resonant saturation, the proposed sequence results in signal fluctuations that are entirely mediated by MT. This time-varying signal is then used for quantitative parameter estimation by applying current MR Fingerprinting (13) (MRF) methods. Images are spatially encoded using an undersampled tiny golden angle radial k-space trajectory (14) and reconstructed using dictionary-based low-rank inversion (LRI) (15) previously used for MRF. The proposed sequence is essentially an MRF acquisition but one which should yield only constant signals if tissues follow the standard Bloch Equations.

We refer to our approach as magnetization transfer-mediated MR fingerprinting (MT-MRF) to reflect these similarities and highlight the fact that induced signal variations are mediated by MT effects. As well as generating quantitative parameter estimates, we also show that the reconstructed time-series data can be used to produce semiquantitative MTR and ihMTR maps. We present an evaluation of the proposed method using phantoms, as well as MT-MRF brain images acquired from five healthy subjects. A comparison with a previous steady-state method is also included (10).

## 2. THEORY

### 2.1. Sequence Description

The proposed sequence consists of a 3D bSSFP acquisition with constant flip angle (FA) and repetition time (TR), in which the excitation pulses are periodically switched between single-band and multiband pulses with different variants as shown in Figure 1. One cycle comprises a number, $n_{\text{MB}}$ of 2-band pulses, followed by a block of $n_{\text{1B}}$ single-band pulses, a block of $n_{\text{MB}}$ 3-band pulses, then another block of $n_{\text{1B}}$ single-band pulses (to allow semisolid saturation to reduce before the next cycle). The on-resonance bands of each pulse are identical, hence a tissue with no semisolid compartment would give constant signal. Tissues with a semisolid component respond transiently as they saturate strongly during multiband periods due to off-resonant bands and recover during single-band periods. Tissues with significant dipolar order effects will also respond differently to the 2-band and 3-band pulses. Example signal curves are shown in Figure 1.

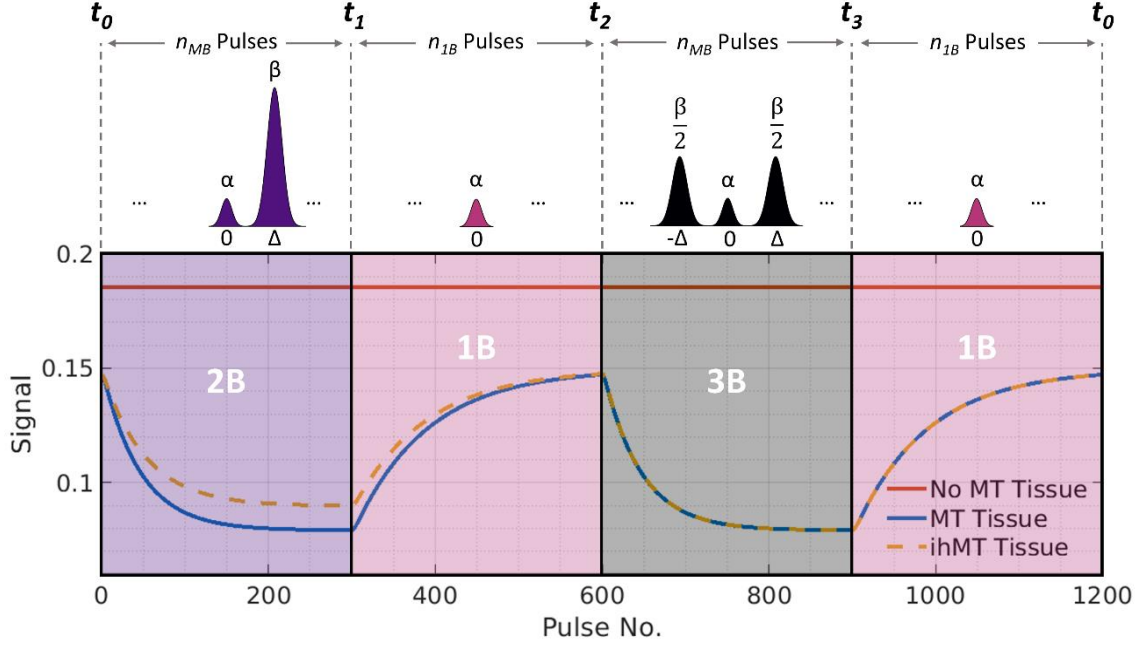

Figure 1: The proposed acquisition uses a rapid gradient echo sequence alternating between the pulses shown in the top row in blocks (bSSFP is used in this work); $n_{MB}$ indicates the number of multiband pulses in a block and $n_{1B}$ is the equivalent number of single-band pulses. The on-resonance part of each pulse (with power α) is identical, so free water (no MT effect) would give constant signal throughout the acquisition. The switched off-resonance bands will affect semisolid proton saturation, resulting in modulation of the free water signal via MT as shown. 2-band (2B) and 3-band (3B) pulses are power-matched (with a combined off-resonance power β) and alternated to give ihMT contrast.

The sequence repeats many times; the total number of excitations in one cycle ($N_{cycle} = 2n_{MB} + 2n_{1B}$) is typically of the order of 1000, such that the cycle repeats after a few seconds for the TRs (~5ms) achieved using bSSFP with multiband pulses. After a small number of repeats, the signal reaches a cyclic-steady-state whose MT or ihMT contrast is dictated by parameters such as $n_{1B}$ and $n_{MB}$ as well as FA, TR, off-resonance power and offset frequency. Qualitative comparison of the efficiency of ihMT contrast generation for different parameter combinations is achieved by defining $\eta$ as peak ihMT contrast generated per square root of time, where TR is the sequence repetition time. Times $t_1$ and $t_3$ are at the ends of 2B and 3B pulse periods, as in Figure 1.

$$\eta = \frac{S(t_1) - S(t_3)}{\sqrt{TR}} = \frac{\Delta ihMT}{\sqrt{TR}} \qquad [1]$$

## 2.2. Spatial Encoding and Image Reconstruction

The cycling of the acquisition sequence can repeat indefinitely and can be made independent of spatial encoding. Since the pulses are spatially non-selective, the encoding must be 3D and cover the entire field-of-view (FOV; the head). Full 3D characterization of each of the $N_{cycle}$ different contrast states is infeasible using regular encoding methods. Instead we take advantage of the fact that changes in signal are compressible in the temporal direction and can be described by a low-rank representation determined from a dictionary of simulations computed for a range of expected tissue properties. As shown by McGivney *et al.* (16), the dictionary **D** can be compressed by singular value decomposition (SVD) and its low-rank approximation can be obtained by multiplication of the dictionary and the matrix of left singular vectors **U**, truncated to a certain rank R to yield $\mathbf{U_R} \in \mathbb{C}^{TNC \times RNC}$ where T is the number of acquired timepoints, N is number of voxels and C is number of coils. The reconstruction problem was previously formulated for MRF as (15–17):

$$\operatorname*{argmin}_{\tilde{\mathbf{x}}, \mathcal{T}_b} \left\lVert \mathbf{U_R FC\tilde{x}} - \mathbf{s} \right\rVert_2^2 + \lambda \sum_b \left\lVert \mathcal{T}_b \right\rVert_* \quad \text{s.t.} \quad \mathcal{T}_b = \mathbf{P_b}(\tilde{\mathbf{x}}) \qquad [2]$$

Here, $\mathbf{s} \in \mathbb{C}^{TKC}$ is the k-space signal where K is k-space trajectory length. In Equation 2, $\mathbf{F} \in \mathbb{C}^{RNC \times RNC}$ performs nonuniform fast Fourier transform and gridding operations; $\mathbf{C} \in \mathbb{C}^{RNC \times RN}$ represents coil sensitivity maps; **x** is a time-series of images sought to be reconstructed and $\tilde{\mathbf{x}}$ is its low-rank approximation ($\mathbf{x} = \mathbf{U_R}\tilde{\mathbf{x}}$ so $\tilde{\mathbf{x}} = \mathbf{U_R^H x}$, where $\mathbf{x} \in \mathbb{C}^{TN}$ and $\tilde{\mathbf{x}} \in \mathbb{C}^{RN}$). Patch-based regularization was added to the problem, as in Bustin *et al.* (18). $\mathbf{P_b}$ constructs a 3D local tensor $\mathcal{T}_b$ around voxel *b* by concatenating local (in each patch), non-local (between similar patches in a neighborhood) and contrast voxels along each dimension (19,20). This problem can be solved using the alternating direction method of multipliers (21); further details can be found in Ref. (18).

There is some flexibility in how much data needs to be collected and how this should be spatially encoded. It has been found that radial or spiral k-space encoding leads to better conditioning of the reconstruction problem than Cartesian sampling (22–24). Furthermore, the framework allows for a flexible amount of undersampling to be used. In total, the amount of data collected would be sufficient to reconstruct $N_V$ fully sampled volumes (ignoring temporal modulation of the signal), while the number of actually reconstructed volumes is R.

## 3. METHODS

### 3.1. Signal Model and Efficient Simulations

The tissue model used in this work (1) consists of a single pool of 'free water' protons, denoted $f$, and a semisolid pool that contains two sub-compartments: one without dipolar order effects, denoted $s1$ (fractional size $1 - \delta$), and the other with dipolar order, denoted $s2$ (fractional size $\delta$). Magnetization is given by vector $\mathbf{M} = \left[M_x^f \ M_y^f \ M_z^f \ M_Z^{s1} M_Z^{s2} \ M_D^{s2}\right]^{\mathrm{T}}$ - subscript Z represents Zeeman ordered longitudinal magnetization and D denotes dipolar ordered longitudinal magnetization. $\mathbf{M}$ evolves as per the "Bloch-McConnell-Provotorov" equations (10,25). These are a set of first-order differential equations summarized in matrix form as $\dot{\mathbf{M}} = \mathbf{AM} + \mathbf{C}$; a complete description using the same notation is in Ref. (10). Key model parameters include: free proton longitudinal and transverse relaxation times, $T_1^f$ and $T_2^f$; semisolid Zeeman and dipolar relaxation times, $T_{1Z}^s = T_{1Z}^{s1} = T_{1Z}^{s2}$ and $T_{1D}^s = T_{1D}^{s2}$; semisolid transverse relaxation time, $T_2^s = T_2^{s1} = T_2^{s2}$; fractional equilibrium magnetizations, $M_0^f$ and $M_0^s$ ($M_0^f + M_0^s = 1$); $f = M_0^s/(M_0^s + M_0^f)$; exchange rate $k$ between free and semisolid pools; and semisolid absorption lineshape $g(\Delta, T_2^s)$, defined for an offset frequency $\Delta$.

Malik *et al.* (10) proposed an efficient means for simulating the steady-state behavior of a sequence of RF pulses and delay periods. The governing equation is written in an 'augmented' form $\dot{\widetilde{\mathbf{M}}} = \widetilde{\mathbf{A}}\widetilde{\mathbf{M}}$, whose solution for a period in which $\widetilde{\mathbf{A}}$ is constant is: $\widetilde{\mathbf{M}}(t + \Delta t) = \exp(\widetilde{\mathbf{A}}(t)\Delta t)\widetilde{\mathbf{M}}(t)$. Evolution over multiple periods can be summarized by a single matrix, which is the product of many matrix exponential terms. The resultant signal can be calculated by finding the eigenvector of this matrix whose eigenvalue is one. Specific details for MT-MRF are given in Supporting Information.

### 3.2. Sequence Design

Sequences of the type shown in Figure 1 were simulated using internal capsule tissue parameters from Mchinda *et al.* (9) and $\eta$ values calculated for different parameter ranges, as shown in Figure 2. Several constraints were considered according to scanner hardware limitations: maximum pulse amplitude, $B_{1,\max} < 20\mu T$, $TR > 2\tau$ (due to RF amplifier duty-cycle limits) and a 'timing' constraint, $TR > 2.5 + \tau$ where 2.5ms is readout duration and $\tau$ is RF pulse duration. Simulated signals were compared to our previous steady-state ihMT ('ss-ihMT') method with scan parameters matching the *in vivo* scheme used in Malik *et al.* (10). The "maximum contrast" solution in Figure 2 was found using a genetic algorithm (*ga* in MATLAB 2019a). The "scan" case was used for all experiments.

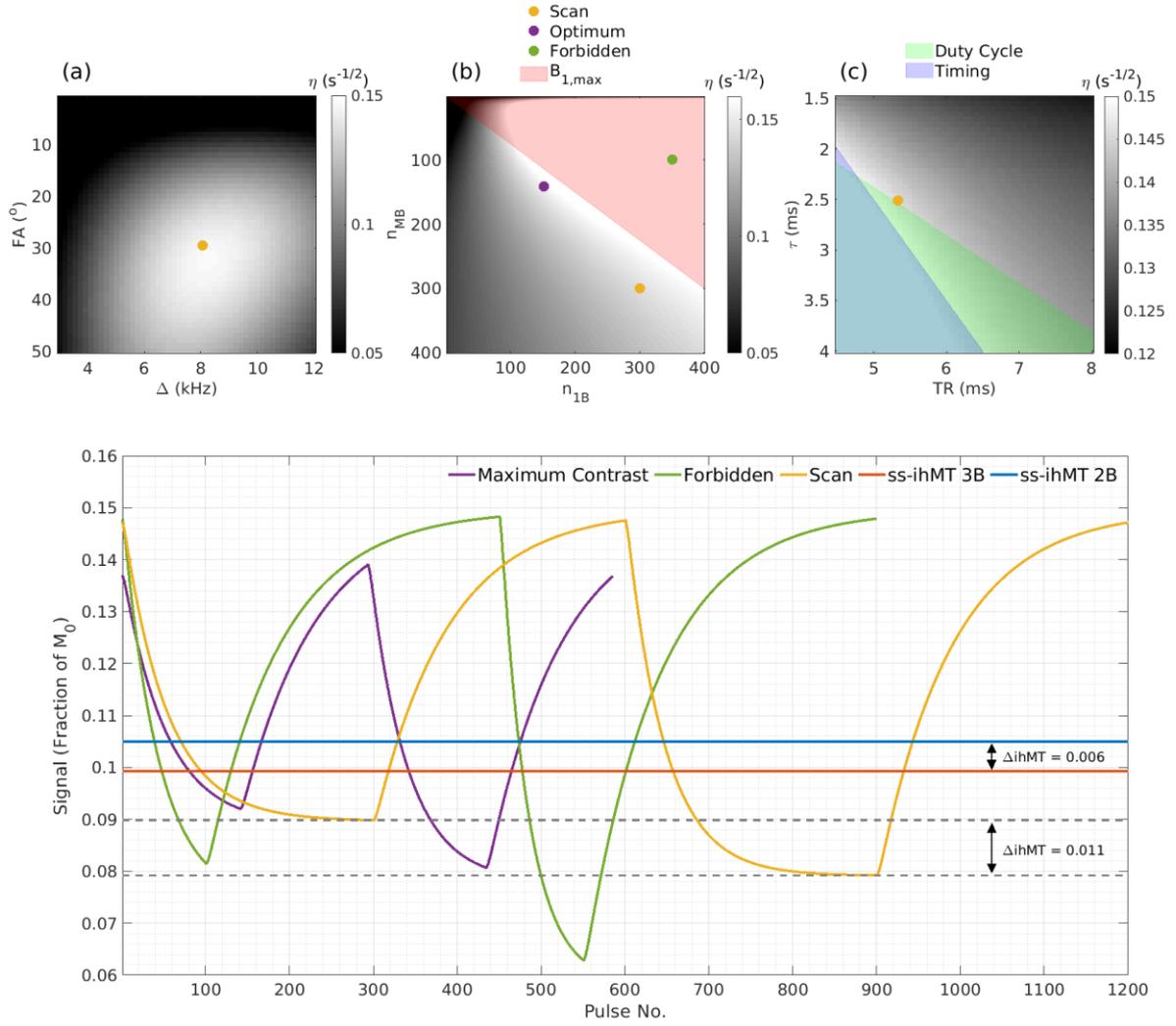

Figure 2: *Top*: $\eta$ as a function of sequence parameters (FA, $\Delta$, $n_{MB}$, $n_{1B}$, $\tau$, TR). In each subfigure, non-plotted parameters are fixed to: TR = 5.3ms, FA = 29.5°, $\tau$ = 2.5ms, $\Delta$ = 8.1kHz, $B_{1,rms}$ = 4µT and $n_{MB} = n_{1B} = 300$. (a) Efficiency as a function of frequency offset and on-resonance flip angle. (b) Efficiency as a function of cycle duration parameters $n_{MB}$ and $n_{1B}$. The colored dots mark exemplar schemes plotted below. 'Scan' parameters are those listed in this caption. (c) Efficiency as a function of TR and pulse duration, showing the operation of timing and duty-cycle constraints. Colored areas are inaccessible parameter combinations that violate hardware constraints. *Bottom*: Signal profiles simulated for the dotted schemes above, and ss-ihMT with equivalent $B_{1,rms}$. The time-variable sequences are ordered 2B-1B-3B-1B (i.e. 2-band pulse period, then 1-band etc.). ihMT contrast is the difference in signals at the end of the 2B and 3B periods. 'Scan' parameters generate greater contrast than the steady-state sequence. The 'maximum contrast' sequence obtains slightly higher contrast, while the 'forbidden' scheme (green) yields far higher contrast but violates peak $B_1$ constraints.

The sequence design for MT-MRF was chosen to give large ihMT contrast, with the constraint that all on-resonance flip angles be constant, i.e. that all signal fluctuations are MT-mediated only. To understand how well this sequence could be used to estimate underlying model parameters, we computed the Cramér-Rao Lower Bound (CRLB) (26–29) and used this to estimate parameter-to-noise ratios (PNRs). Since the PNR and image signal-to-noise ratio (SNR) are both dependent on the underlying acquisition noise level, we have quoted the ratio PNR/SNR that independently quantifies the precision relative to image SNR. Figure 3 shows PNR/SNR when estimating different combinations of parameters with others fixed at their true values (fixed parameters shown as white). For all further analysis, white matter (WM)/gray matter (GM)/GM-WM average parameter sets from Varma *et al*. (12) are considered and correspond to respective values: $R_1^f$ (i.e. $1/T_1^f$) = 0.92/0.55/0.735s$^{-1}$, $T_{1Z}^s$ = 1/1/1s, $T_2^f$ = 69/99/84ms, $T_{1D}^s$ = 6.2/5.9/6.05ms, $T_2^s$ = 9/7.58/8.28μs, $k$ = 59.6/50.8/55.2s$^{-1}$ and $f$ = 0.101/0.0348/0.0679. $\delta$ = 1 (i.e. a single semisolid compartment model) is used to simplify parameter estimation as per recent quantitative ihMT studies. (10,12) From Figure 3 it appears that $T_1^f$, $f$ and $T_{1D}^s$ can be estimated to reasonable precision with others held fixed (Combination 2), and this motivates the approach taken in subsequent sections.

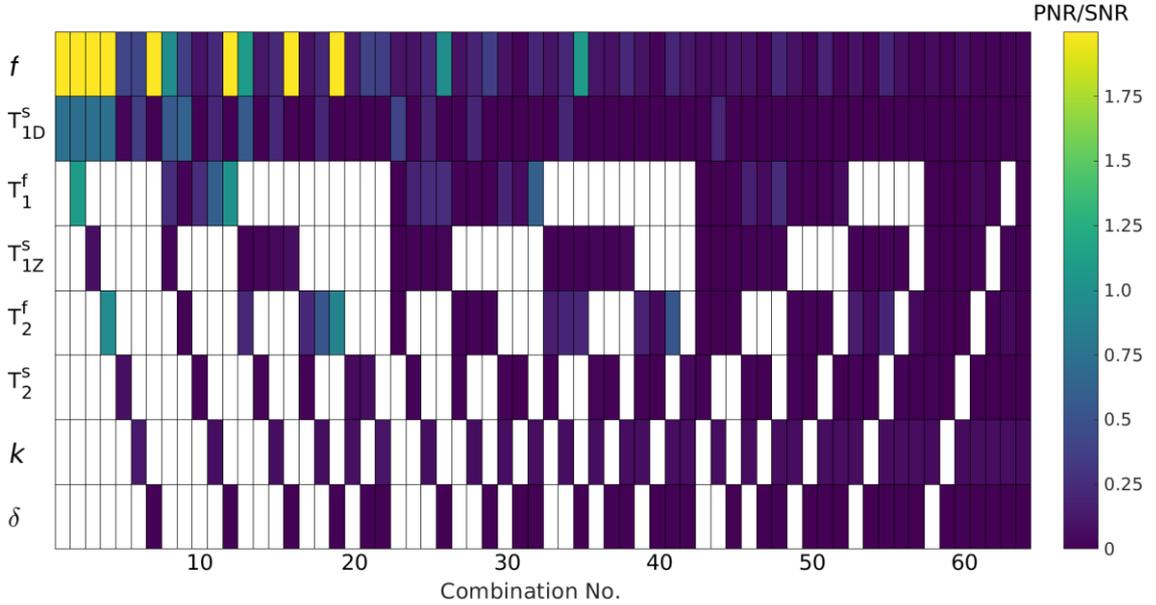

Figure 3: Calculation of PNR/SNR for different parameter combinations. Colored squares indicate parameters that are estimated for a given combination (column), whereas white squares indicate fixed parameters (i.e. the left-most columns feature the most fixed parameters and the right-most columns feature the most estimated parameters). To achieve a reasonable PNR, only three parameters can be estimated simultaneously. Combination 2 is used for all dictionary fits.

## 3.3. Dictionary Generation, Low-Rank Basis and Parameter Estimation

Two types of dictionary were required for this work. Firstly, a 'full' dictionary denoted **D** included variation of all model parameters. Having nine free parameters and a long temporal duration of each atom (1200 readouts) meant that the dictionary would quickly become too large for the memory of the PC used for calculation (8(16) × Intel® Core™ i7-5960X 3.00 GHz CPU, 64 GB RAM). Hence a version of **D** was constructed using coarse sampling with the sole purpose being to discover a lower rank basis as detailed above. Secondly for parameter estimation, a reduced dictionary **D̃** was constructed using finer sampling of the three estimated parameters (with all others fixed).

**D** was created with ∼1 million atoms using parameter value ranges: $T_1^f = 0.2: 0.5: 3.7\text{s}$, $T_{1Z}^s = 0.1: 0.2: 0.9\text{s}$, $T_{1D}^s = 2: 4: 22\text{ms}$, $T_2^f = 50: 70: 470\text{ms}$, $T_2^s = 5: 3: 20\mu\text{s}$, $\delta = 0: 0.2: 1$, $k = 40: 20: 100\text{s}^{-1}$, $f = 0: 0.05: 0.25$ and $B_0$-induced phase per TR $-\pi: \frac{\pi}{3}: \pi$ since the method is based on bSSFP. The semisolid lineshape was assumed to be super-Lorentzian (30).

The low-rank basis $\mathbf{U_R}$ was constructed by performing an SVD on **D**. A further reduction by randomly sampling 600,000 atoms was needed to perform the SVD; the random sampling was done ensuring an equal number of "no semisolid" ($f = 0$), "no dipolar order" ($f \neq 0, \delta = 0$) and "ihMT" entries ($f \neq 0, \delta \neq 0$). The ability of the resulting basis to reproduce arbitrary tissue signals was assessed by projecting some test signals into the low-rank space via multiplication by $\mathbf{U_R^H}$ (truncated to different ranks) then back into the time domain by $\mathbf{U_R}$. These approximated signals were compared to their ground-truth simulated equivalents (Figure 4). Test signals were not themselves dictionary atoms but represented expected tissue values for cerebrospinal fluid (CSF), GM and WM. CSF was assigned: $T_1^f = 3\text{s}$, $T_2^f = 2\text{s}$, and $f = 0$, whilst GM/WM took their respective values from Section 3.2.

Reduced dictionary **D̃** was generated to estimate only $T_1^f, f$ and $T_{1D}^s$ simultaneously using 100 increments of each parameter over their feasible ranges. The size of this dictionary was further reduced by using the low-rank basis discovered from **D** such that more increments could have been used if required; it was found that exceeding 100 increments for each parameter was unnecessary, with changes below the noise level when trialed on the acquired data. Unless stated otherwise, all reduced dictionaries used $T_{1Z}^s = 1\text{s}$, $T_2^f = 84\text{ms}$ and $T_2^s = 8.28\mu\text{s}$ (GM-WM average in Section 3.2).

In addition to parameter estimation, reconstructed data (**x̃**) can be projected back to the time domain (**x**) from which familiar semiquantitative metrics, MTR and ihMTR, can be calculated. The degree of contrast varies through time (since the signals are time-varying), hence we define these metrics using the minimum and maximum signal values:

$$\text{MTR} = 1 - \frac{S(t_1) + S(t_3)}{S(t_0) + S(t_2)} \qquad [3]$$

$$\text{ihMTR} = 2\frac{S(t_1) - S(t_3)}{S(t_0) + S(t_2)} \qquad [4]$$

where $t_0$ to $t_3$ (as indicated in Figure 1) are times within the cycle.

### 3.4. Imaging Experiments

All experiments were performed on a 1.5T Philips Ingenia MRI system with a 15-channel head coil for signal reception.

### 3.4.1. System Characterization and Correction for RF Instability

Our sequence was first characterized using 1D phantom scans (test-tube phantom with no phase encoding), allowing direct measurement of the full temporal signal with no need for low-rank reconstruction. Inconsistencies were found between the on-resonance amplitudes and phases of each RF pulse type, causing oscillations in signal when switching between them. The origin of these inconsistencies could not be determined but were stable for a given pulse configuration. Empirically determined complex scaling factors for each pulse were used to ensure the on-resonance component of each pulse type was the same. See Supporting Information for more details.

### 3.4.2. Sequence Implementations

For MT-MRF, a 3D tiny golden angle radial, "stack-of-stars" k-space trajectory with one spoke per RF pulse was used for all experiments (14), where TR = 5.3ms, TE = 2.7ms and FA = 29.5˚ (additional sequence details below). Encoding was ordered so as to loop over repeats of each radial 'star', before then looping over $k_z$ partitions. Although not necessary for the method to work, the total number of radial spokes acquired per partition was set to be commensurate with $N_{\text{cycle}}$ so that the same $k_x$-$k_y$ sampling was used for all $k_z$ partitions, enabling independent reconstruction of each z-location. As a reference, ss-ihMT (10) was also used, in this case with Cartesian k-space sampling, acquiring separate images using 1-band, 2-band and 3-band RF pulses.

### 3.4.3. Phantom Validation

MT-MRF data were acquired on three phantoms: MnCl$_2$-doped water (0.05mM), cross-linked bovine serum albumin (BSA, 10% w/w in water prepared as in Koenig *et al.* (31)) and prolipid 161 ('PL161', 15% w/w; Ashland Inc, Covington, Kentucky USA; prepared as in Swanson *et al.* (2) but excluding any T$_1$ reducing agent). These phantoms are examples of "no semisolid", "no dipolar order", and "ihMT" materials respectively. FOV of $115 \times 115 \times 150$mm and isotropic resolution 1.5mm³ were used with a total acquisition time of 21m32s. Quantitative maps of $T_1^f, f$ and $T_{1D}^s$ were computed using the reduced dictionary. In practice, it was observed that the large spread of $T_2^f$ and $T_1^f$ between phantoms led to inconsistent results, and so a second fit was performed using an alternative dictionary with the same fixed parameters except $T_2^f = 130$ms. Quantitative parameter estimates were compared against ss-ihMT results from the same phantoms, acquired as part of a previous study (10). That study used the same resolution but acquired data over multiple excitation flip angles with a significantly longer total scan duration of 60m48s. Quantitative maps were produced by dictionary matching using separate dictionaries relevant to this sequence.

### 3.4.4. *In vivo* Study

Human scanning was performed on 5 healthy subjects (3 males and 2 females, age 29±7.5 years) who gave prior written consent in line with local ethical approval. MT-MRF experiments used FOV = $250 \times 250 \times 180$mm and 1.5mm³ isotropic resolution with sagittal orientation to avoid fold-over artefacts. 2400 spokes ($2 \times N_{\text{cycle}}$) were acquired per slice encode position, equating to $N_V \sim 9$ and a scan time of 25m48s. One subject was scanned twice using MT-MRF (eight months apart). Quantitative maps of $T_1^f, f$ and $T_{1D}^s$ were computed using the reduced dictionary.

For comparison, ss-ihMT datasets were also acquired for every subject. These acquisitions used the same resolution, FOV, TR, TE and FA as MT-MRF scans; 6 signal averages were used to match acquisition duration (each average = 1m23s). For ss-ihMT data, motion correction using rigid body registration (*imregister* in MATLAB 2019a) was run *before* signal averaging and subsequent calculation of MTR/ihMTR as in (10). Spatial smoothing with Gaussian kernels (standard deviation 1.05mm for MTR and 1.35mm for ihMTR) was used to improve SNR. Datasets for each subject were aligned and regions-of-interest (ROIs) drawn in corticospinal tracts (CST), frontal WM, and cortical GM for numerical analysis. ss-ihMT can produce MTR/ihMTR but no other parameter estimates.

## 4. RESULTS

### 4.1. Simulation of Obtainable Contrast

Figure 2 shows ihMT contrast efficiency ($\eta$) calculated for a range of acquisition parameters. The colored areas are excluded since they violate constraints on peak $B_1$, duty-cycle or timing. A clear region of maximum $\eta$ exists between FA = 25-40° and $\Delta$ = 7-9kHz. Given the optimal FA and $\Delta$, Figure 2b shows some example sequences with different $n_{1B}$ and $n_{MB}$ in a single cycle, marked by colored dots. The purple scheme has maximal contrast, though the dot marked 'scan' (yellow) was used for our experiments, with sequence parameters: TR = 5.3ms, FA = 29.5°, $\tau$ = 2.5ms, $\Delta$ = 8.1kHz, $n_{MB} = n_{1B} = 300$ and a constant $B_{1,rms} = 4\mu T$ (maximum achievable). This scheme was selected because it yielded more gradual contrast variations than the 'maximum contrast' case and is expected to provide favorable properties for reconstruction (24,32). A 'forbidden' scheme was also simulated (green) that exceeds maximum $B_1$ constraints but generates even greater ihMT contrast ($\Delta$ihMT).

Simulated signal profiles are also compared in Figure 2 according to the protocols highlighted in the above heat maps, and with an ss-ihMT scan with equal time-averaged RF power ($B_{1,rms}$). One cycle of each sequence is shown using ordering 2B-1B-3B-1B. $\Delta$ihMT is found by comparing the end of the 2B period with the end of the 3B period ($t_1$ and $t_3$, as labelled in Figure 1). The 'scan' sequence generates almost twice the contrast achieved using ss-ihMT. The respective efficiencies of contrast generation are $\eta = 0.080s^{-1}$ and $\eta = 0.146s^{-1}$. The 'maximum contrast' sequence yields only a marginally (~6%) greater ihMT effect, though the 'forbidden' scheme contrast is significantly higher.

### 4.2. Dictionary Analysis

Figure 4 displays the singular vectors (Figure 4a) and corresponding singular values (Figure 4b) of a dictionary corresponding to the 'scan' sequence. The first five components describe 99.2% of the 'energy' within the dictionary. Figure 4c plots the mean residuals in representing WM, GM, and CSF 'test' signals for different R. Improved representation of each tissue type occurs for higher R as expected but by R = 5, all three tissue types are accurately recreated with negligible residuals.

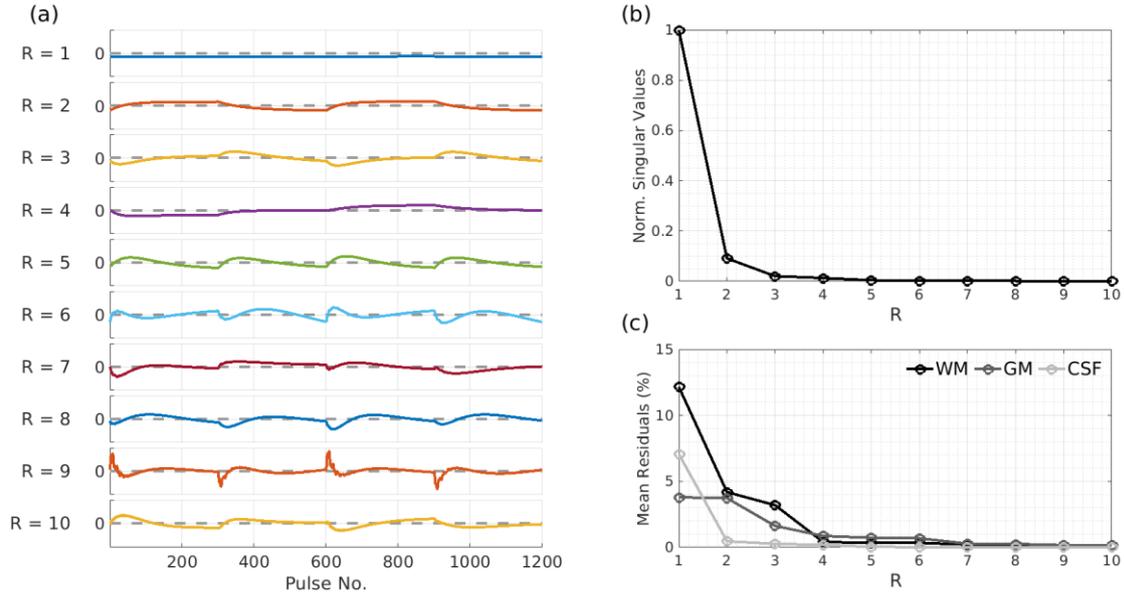

Figure 4: Low-rank analysis of the dictionary **D**. (a) The first ten singular vectors of the dictionary; each axis is scaled identically to a [-0.2-0.2] range. Each subplot represents one column of $U_R$ and dashed lines indicate the position of zero along each y-axis. (b) Singular values corresponding to each singular vector, normalized to the first singular value. (c) Mean residuals between simulated WM, GM and CSF signal curves and equivalent low-rank approximations for different R. The three plausible MT-MRF signals are better represented as R increases and are sufficiently described by R = 5.

### 4.3. Image Reconstructions and Semiquantitative Maps

Figure 5a displays a single slice from the MT-MRF phantom acquisition in the reduced basis (i.e. $\tilde{\mathbf{x}}$); note the changes in intensity scale as R increases. Figure 5b shows example data projected back into the time domain, with MTR and ihMTR maps computed from these in Figure 5c. The MTR map shows non-zero values in BSA and PL161 and no effect in $MnCl_2$-doped water, as expected. The ihMTR map shows a significant effect only in PL161, again as expected. ROI measurements in the reconstructed images agree with direct non-phase-encoded (NPE; 1D) measurements in Figure 5d.

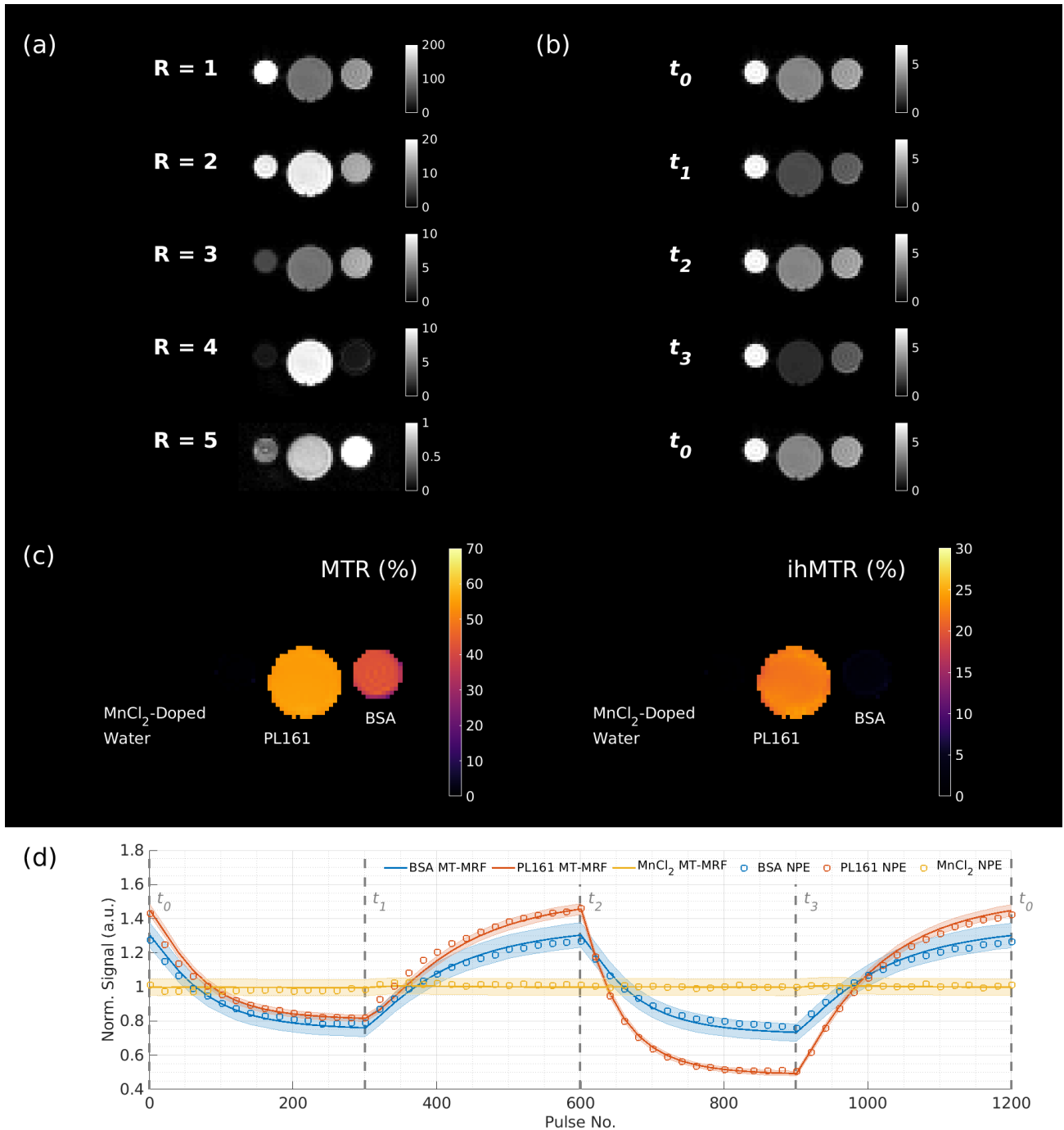

Figure 5: (a) Example single-slice images from the low-rank reconstructed phantom dataset ($\tilde{\mathbf{x}}$). Locations of the three tubes are as labelled in (c). (b) The five displayed components are then transformed into the time domain to yield 1200 volumes of data; one slice is shown for the key timepoints labelled in Figure 1 ($\mathbf{x}$). (c) Corresponding MTR and ihMTR maps show the expected contrast in each tube: ihMT effects are only seen in PL161 and MT effects, only in BSA and PL161; $MnCl_2$-doped water has MTR = ihMTR = 0. (d) Signal time-courses from central regions-of-interest in each phantom, compared to results from a non-phase-encoded experiment (NPE); the signal is faithfully represented by the low-rank reconstruction (these data are not fitted to one another, only over-plotted).

Figure 6 shows MTR and ihMTR maps of the brain from all five subjects. Table 1 and Supporting Information Table S2 compare these metrics between each method for matched ROIs after image registration. Supporting Information Figure S4 is a test-retest comparison for the same subject, scanned eight months apart.

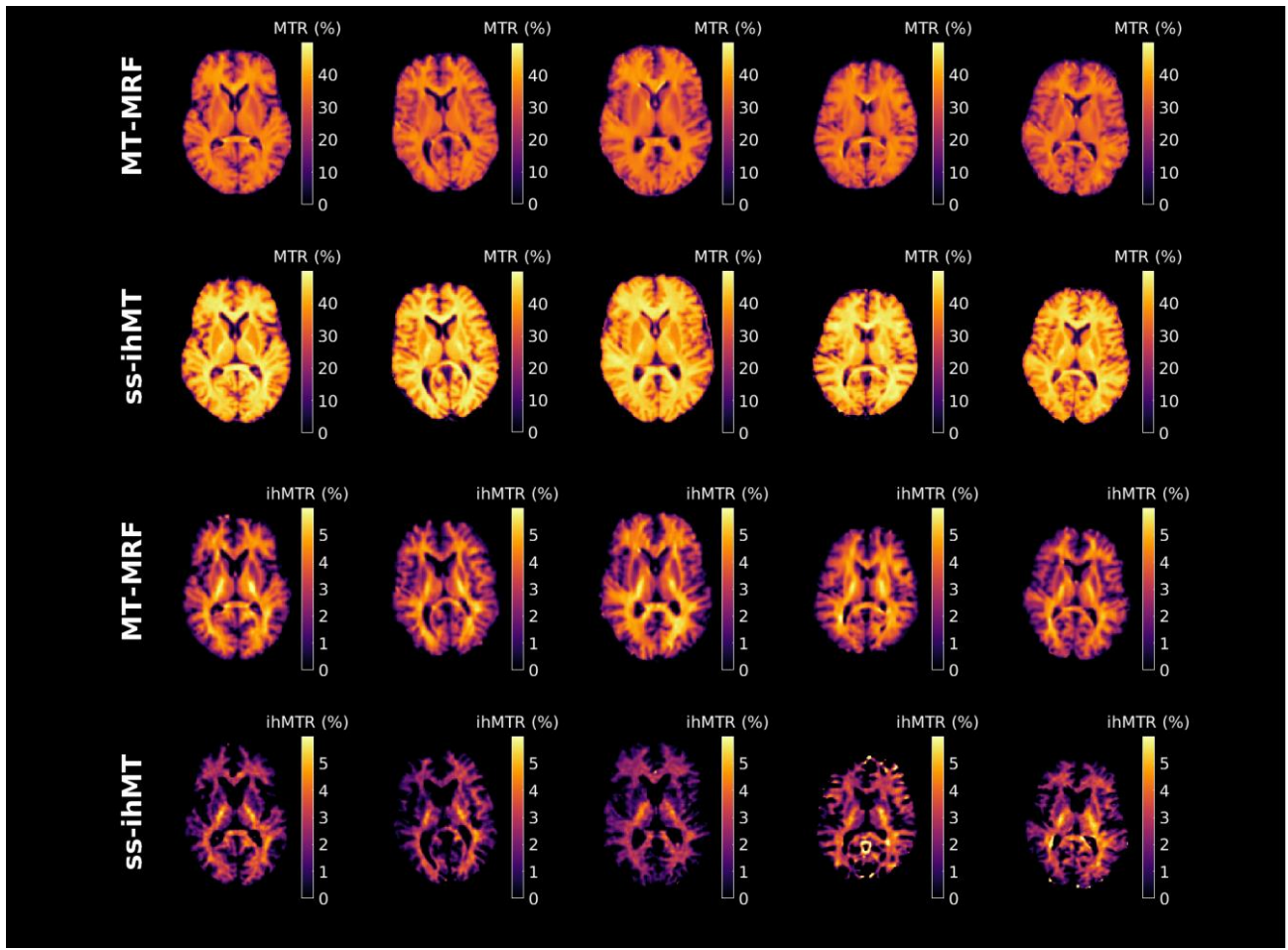

Figure 6: Example single-slice *in vivo* MTR and ihMTR maps from ss-ihMT and MT-MRF and each subject (separate columns). The former shows strong GM-WM contrast, but the latter consistently appears more correlated with WM.

Table 1: *Top*: ihMTR values obtained from different regions-of-interest: frontal white matter, cortical gray matter, and corticospinal tracts in the axial slices from five subjects (1-5; left-to-right) in Figure 6. *Bottom*: MT-MRF quantitative parameter estimates obtained for the same axial regions-of-interest from the five subjects in Figure 7.

| ihMTR (%) | ss-ihMT | | | MT-MRF | | |
|---|---|---|---|---|---|---|
| | CST | Frontal WM | Cortical GM | CST | Frontal WM | Cortical GM |
| **Subject 1** | 4.96 ± 0.41 | 2.91 ± 0.41 | 2.06 ± 0.18 | 5.71 ± 0.17 | 4.51 ± 0.19 | 2.96 ± 0.35 |
| **Subject 2** | 4.58 ± 0.54 | 3.11 ± 0.50 | 1.82 ± 0.12 | 5.52 ± 0.18 | 4.66 ± 0.11 | 2.88 ± 0.24 |
| **Subject 3** | 4.10 ± 0.54 | 2.57 ± 0.37 | 1.71 ± 0.47 | 5.69 ± 0.19 | 4.52 ± 0.31 | 2.87 ± 0.23 |
| **Subject 4** | 4.44 ± 0.40 | 3.08 ± 0.37 | 2.51 ± 0.45 | 5.89 ± 0.18 | 4.73 ± 0.21 | 2.67 ± 0.37 |
| **Subject 5** | 4.19 ± 0.64 | 2.63 ± 0.54 | 2.12 ± 0.37 | 5.31 ± 0.15 | 4.65 ± 0.37 | 2.99 ± 0.34 |
| | $f$ | | | $T_{1D}^s$ (ms) | | |
| | CST | Frontal WM | Cortical GM | CST | Frontal WM | Cortical GM |
| **Subject 1** | 0.095 ± 0.003 | 0.097 ± 0.006 | 0.081 ± 0.004 | 6.22 ± 0.12 | 3.64 ± 0.34 | 2.66 ± 0.35 |
| **Subject 2** | 0.094 ± 0.001 | 0.094 ± 0.005 | 0.070 ± 0.003 | 5.55 ± 0.14 | 3.64 ± 0.41 | 2.64 ± 0.23 |
| **Subject 3** | 0.093 ± 0.002 | 0.095 ± 0.004 | 0.075 ± 0.003 | 5.94 ± 0.13 | 3.76 ± 0.26 | 2.73 ± 0.25 |
| **Subject 4** | 0.099 ± 0.002 | 0.091 ± 0.003 | 0.076 ± 0.003 | 5.48 ± 0.13 | 3.71 ± 0.22 | 2.64 ± 0.28 |
| **Subject 5** | 0.092 ± 0.001 | 0.089 ± 0.005 | 0.074 ± 0.004 | 5.24 ± 0.17 | 3.75 ± 0.25 | 2.64 ± 0.30 |

## 4.4. Quantitative Parameter Estimation

Figures 7- 8 show example parameter maps for semisolid fraction $f$, dipolar relaxation time $T_{1D}^s$ and $T_1^f$. The phantom study (Table 2) compares values from MT-MRF against fitting to ss-ihMT data using multiple flip angles (details in Ref. (10)). Good agreement is observed between $f$ and $T_{1D}^s$ estimates, with MT-MRF yielding tighter distributions (lower variance) even though the ss-ihMT data were acquired over almost three times longer examination. Two sets of fixed parameters were used to estimate phantom properties ($T_2^f$ = 84ms and $T_2^f$ = 130ms; all others equal). Estimates of $f$ and $T_{1D}^s$ are less sensitive to this change and agree within precision when using either fixed parameter set. Though for BSA and PL161, $T_1^f$ changes significantly when fixed $T_2^f$ is altered. Corresponding MT-MRF parameter maps for these two scenarios are shown in Supporting Information Figure S3.

Table 2: Comparison of phantom dictionary fitting results for MT-MRF and ss-ihMT. Two independent fits were performed: one with $T_2^f$ more aligned to BSA (84ms) and another with $T_2^f$ more aligned to PL161 (130ms).

| | | $T_2^f$ = 84ms | | $T_2^f$ = 130ms | |
|---|---|---|---|---|---|
| | | **MT-MRF** | **ss-ihMT** | **MT-MRF** | **ss-ihMT** |
| **BSA** | $T_1^f$ (s) | 1.35 ± 0.08 | 1.43 ± 0.30 | 1.05 ± 0.04 | 1.83 ± 0.38 |
| | $f$ | 0.081 ± 0.005 | 0.087 ± 0.009 | 0.082 ± 0.005 | 0.075 ± 0.009 |
| | $T_{1D}^s$ (ms) | 0.44 ± 0.38 | 1.3 ± 3.0 | 0.42 ± 0.39 | 2.7 ± 3.6 |
| **PL161** | $T_1^f$ (s) | 2.02 ± 0.15 | 1.12 ± 0.15 | 1.40 ± 0.10 | 1.41 ± 0.18 |
| | $f$ | 0.177 ± 0.008 | 0.188 ± 0.014 | 0.175 ± 0.006 | 0.168 ± 0.014 |
| | $T_{1D}^s$ (ms) | 24.6 ± 0.9 | 26 ± 4.0 | 24.5 ± 1.0 | 28.4 ± 4.1 |

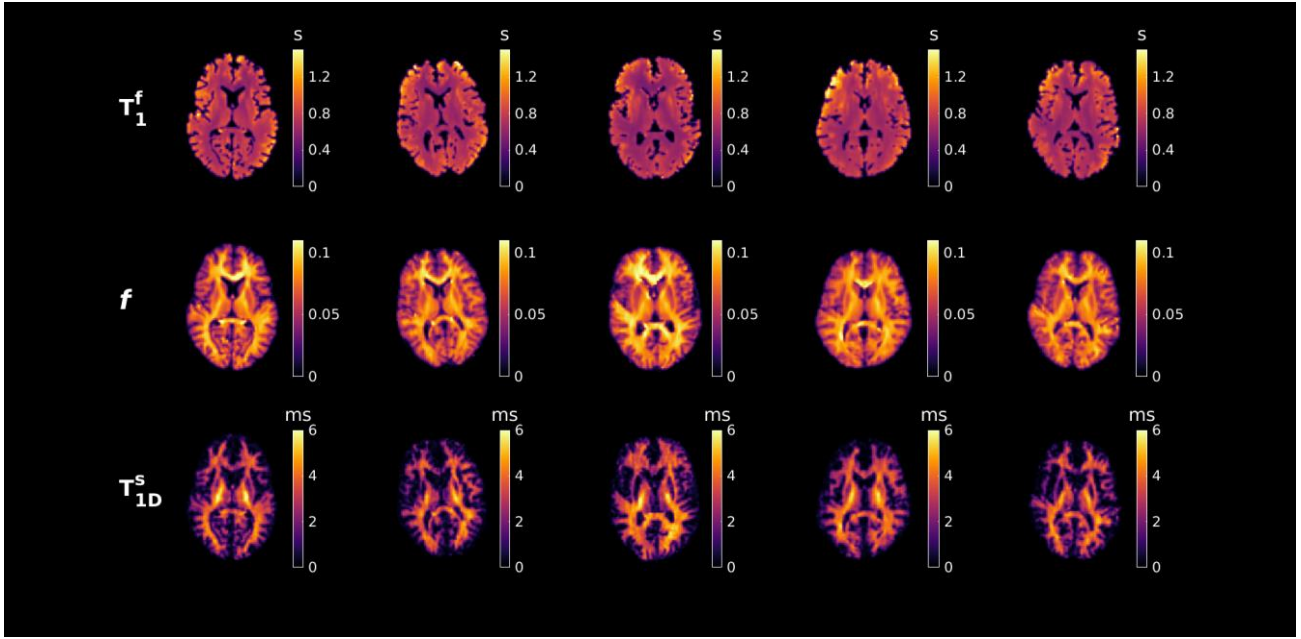

Figure 7: Estimation of free pool $T_1^f$, semisolid fraction ($f$) and dipolar $T_{1D}^s$ from a dictionary fit to *in vivo* data from each subject (separate columns). Maps for $f$ show strong GM-WM contrast but CST are more discernible in $T_{1D}^s$ maps.

*In vivo* parameter maps from Subject 1 are shown in Figure 8 for single axial, coronal and sagittal slices. Figure 7 presents consistent parameter maps across all subjects, with $f$ and $T_{1D}^s$ showing apparently distinct contrasts. To demonstrate this, a joint-histogram of $f$ and $T_{1D}^s$ in GM and WM pixels of Subject 1 (masks created using FSL5.0 BET and FAST (33)) is included in Figure 8.

**PRE-PRINT: SUBMITTED TO MAGNETIC RESONANCE IN MEDICINE**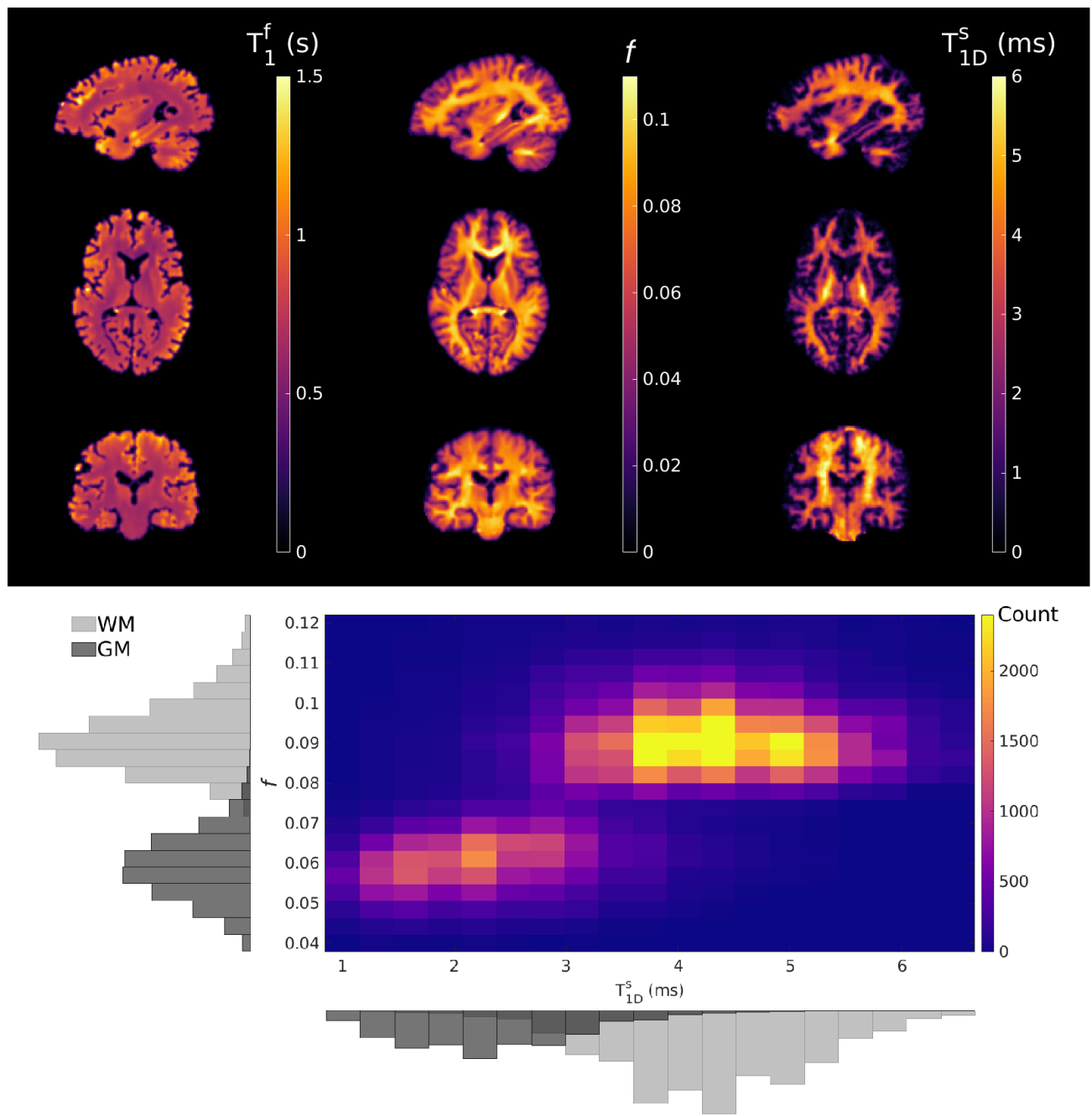

Figure 8: *Top*: Example parameter maps obtained from a dictionary fit to *in vivo* data from Subject 1 (male, aged 41) in three orientations. *Bottom*: Joint-histograms of $f$ and $T^s_{1D}$ produced from 3D GM and WM masks. These two parameters show different contrasts and although strongly correlated, their relationship is not directly proportional, suggesting complementary information is provided.

Both semiquantitative metrics (Figure 6, Table 1, and Supporting Information Table S2) and quantitative metrics (Figure 7, Table 1) show inter-subject and test-retest (Supporting Information Figure S4) repeatability, with MT-MRF yielding a higher ihMTR across all chosen brain regions. Supporting Information 5 presents a series of alternative fits and brief bias investigation as fixed $T_{1Z}^s$, $T_2^f$, $T_2^s$ and $k$ values are changed. Since tissues with $f \sim 0$ provide no MT-mediated contrast, $T_1^f$ cannot be estimated for these - this is the case for CSF and MnCl$_2$-doped water. When presenting these results, voxels in the $T_1^f$ map with $f < 0.04$ are set to zero; the threshold here was determined by examining Figure 8, which shows that all GM and WM pixels are expected to exceed this value.

## 5. DISCUSSION

This work uses a balanced SSFP sequence with multiband RF pulses that are periodically switched to allow variable off-resonant saturation (i.e. none, single frequency, and dual frequency) while keeping the on-resonance flip angle constant. MT-MRF never directly breaks the steady-state of the water magnetization but MT-mediated signal fluctuations of around 50% are observed *in vivo*. These transient signal changes are reconstructed using techniques proposed for MR fingerprinting. Indeed, as presented, the sequence is essentially a very simple MRF acquisition that would yield constant signals (and hence be unable to perform any parameter estimation) in the absence of MT. The resulting measurements are used for quantitative estimation of some of the underlying model parameters via dictionary matching but also to reconstruct semiquantitative MT/ihMT ratio maps.

**MT-MRF for Semiquantitative Mapping**

Though reconstruction of artefact-free time domain images is not usually associated with classic MRF methods, the low-rank inversion (15) approach does produce high-quality 'singular' images ($\tilde{\mathbf{x}}$) that would normally be passed directly to dictionary matching. It is also possible to map these back to the time domain ($\mathbf{x}$) and we have used this property to produce MTR/ihMTR maps using points of maximum contrast within the time-evolving signals as references (Equations 3-4).

The initial motivation for developing MT-MRF was to improve the contrast achieved from the previously developed ss-ihMT method (10). While ss-ihMT used separate image acquisitions with and without off-resonant saturation, it was shown by Varma *et al.* (12) that a 'low duty-cycle' scheme, alternating between high and low saturation power, would boost contrast. Sequence parameters were optimized with this in mind, trying to maximize the peak ihMT contrast (normalized to $\sqrt{\text{TR}}$). The extent of this improvement is shown in Figure 2 - ihMT contrast is almost doubled versus ss-ihMT.

Figure 5 shows the reconstructed images from a phantom experiment in both the reduced basis ($\tilde{\mathbf{x}}$) and time domain ($\mathbf{x}$), including a comparison of reconstructed time domain signals with non-phase-encoded measurements (Figure 5d). The low-rank inversion reconstruction (using R = 5) is able to faithfully represent the time domain signal. The resulting MTR map is non-zero in PL161 and BSA phantoms, while the ihMTR map is non-zero only for PL161, as anticipated (10).

Figure 6 shows MT-MRF MTR and ihMTR maps from a study on five healthy subjects, compared to those from ss-ihMT. Contrast is consistent across subjects (Tables 1 and S2), whilst a test-retest validation on one subject (Supporting Information Figure S4; Table S3) also showed very consistent contrast. MT-MRF generally finds greater ihMTR than for ss-ihMT (62% greater in frontal WM, for example) but lower MTR; since the scheme was optimized for ihMT, the signal never fully returns to equilibrium. Standard deviations in Table 1 are smaller for MT-MRF. This may be due to a combination of better CNR, superior motion robustness since a radial readout was used, and use of spatial regularization (Equation 2).

**Quantitative Parameter Estimation**

The generation of MTR and ihMTR maps is a limited use of the time-series data available; quantitative parameter estimation is also possible. CRLB calculation (Figure 3) suggested that not all parameters can be estimated to good precision, thus it was decided to only reconstruct $T_1^f$, $f$ and $T_{1D}^s$. Phantom experiments using MT-MRF were compared to previous measurements using ss-ihMT with variable flip angles for estimation of these three parameters (Table 2). Estimates of $f$ and $T_{1D}^s$ agree for both methods, and the MT-MRF sequence provides better precision, particularly for $T_{1D}^s$. The estimated $T_{1D}^s \sim$ 25ms for PL161 is in line with measurements from Swanson et al (2). Fits using different $T_2^f$ were performed because BSA and PL161 have very different properties (more so than brain tissue) and thus fixing parameters was found to cause different $T_1^f$ estimation bias for MT-MRF and ss-ihMT. However, MT parameters estimates (i.e. $f$ and $T_{1D}^s$) were consistent regardless of $T_2^f$.

*In vivo* results in Figure 7 show consistent parameter values across subjects. $f$ and $T_{1D}^s$ are significantly different in GM compared to WM and $T_{1D}^s$ seems to highlight strongly myelinated structures (e.g. CST). Generally, parameter estimates between different subjects are in agreement (Table 1), with the highest $T_{1D}^s$ values found in CST (~5.5ms) and lower values (~2.2ms) found in cortical GM; orientation with respect to $B_0$ may influence these values (34). Estimates for $f$ exhibit higher precision - an observation that is supported by the CRLB analysis in Figure 3. Maps of $f$ and $T_{1D}^s$ do show distinct contrasts; Figure 8 presents a whole-brain joint-histogram from one subject and demonstrates that these parameters are correlated but not simply linearly dependent.

Compared to the literature, CST $T_{1D}^s$ estimates (~6.0ms in the coronal slice in Figure 8) are generally in agreement with those reported elsewhere (35) and WM estimates for $f$ concur with those in Varma *et al.* (12). Others (35,36) have reported GM $T_{1D}^s$ to be similar to WM (~5.6ms), however Swanson *et al.* also found far shorter $T_{1D}^s$ in GM compared to WM from bovine spinal cord. (2)

**Parameter Estimation Biases**

Parameter fixing is common in quantitative MT imaging because model parameters can influence the signals in similar ways, so estimation is not well-posed and may create bias. $T_1^f$, $f$ and $T_{1D}^s$ are estimated here, whilst $T_{1Z}^s$, $T_2^f$, $T_2^s$ and $k$ are fixed. Similar methods have been used by others: Varma *et al.* fixed $k$, $\frac{kf}{T_1^f}$, $\frac{1}{T_1^f T_2^f}$ and $T_2^s$ to estimate $T_{1D}^s$; Hilbert *et al.* (37) fixed $T_1^f = T_{1Z}^s$, $k$ and $T_2^s$ to estimate $T_1^f$, $T_2^f$ and $f$ using MRF; and Cohen *et al.* (38) fixed $T_1^f$, $T_{1Z}^s$ and $T_2^s$ in a CEST-MRF study.

In our work, $f$ and $T_{1D}^s$ estimates are comparable to those found by others. The phantom experiment provided similar estimates from two different experiments, though separate fixed values of $T_2^f$ were considered. The main impact of changing $T_2^f$ is to alter the estimate of $T_1^f$, which could indicate that this is acting as a 'nuisance' parameter to absorb uncertainties in the relaxation times. Supporting Information Figure S5c explores this further for *in vivo* data by examining the effects of changing each fixed model parameter by ±10%. It is observed that $T_{1Z}^s$ has negligible impact, while changing $k$ causes small changes in $f$ and $T_{1D}^s$. In line with phantom experiments, changing $T_2^f$ causes different $T_1^f$ estimates but has negligible impact on $f$ and $T_{1D}^s$. The most significant correlation exists between $T_{1D}^s$ and $T_2^s$, which has also been reported by Varma *et al.* (35).

Recent work from Wang *et al.* (39) has questioned the commonly made assumptions that $T_{1Z}^s$ can either be fixed at some simple value (i.e. 1s) or set equal to $T_1^f$ (40,41). Instead, Wang *et al.* suggest that $T_{1Z}^s$ is a major determinant of observed T₁ and is strongly frequency dependent; a $T_{1Z}^s$ of approximately 120ms in WM should be expected at 1.5T. Figure 8 used the assumption $T_{1Z}^s = 1s$ but we repeated *in vivo* dictionary fitting for Subject 1 using a fixed value of 200ms; the result in Supporting Information Figure S5a shows substantially longer $T_1^f$ but largely unchanged $f$ and $T_{1D}^s$. CRLB analysis (Figure 3) showed that $T_2^f$ could be estimated instead of $T_1^f$ - this was also performed (fixing $T_1^f = 1.36s$ from Section 3.2) and the result (Supporting Information Figure S5b) shows that a plausible T₂ map can be obtained but again, there is little impact on estimated $f$ and $T_{1D}^s$.

**Determining Subspace Rank**

The embodiment of MT-MRF used in this work was chosen to emulate a 'standard' MT measurement that cycles over different saturation settings with the same flip angle; it has the interesting property that all signal variation is MT-mediated. One important side-effect of this is that parameter estimation is not possible for substances without a semisolid compartment. However, the fairly simple time-structure of this specific sequence leads to smooth temporal variations such that the generated dictionaries have an effective rank of five. Time domain signals from representative tissues could be reconstructed with <0.75% error using the reduced basis. The presented images used rather long acquisitions (~25 minutes) which were necessary because ihMT contrast is rather weak. In one sense, the images are highly undersampled since approximately nine complete volumes of k-space data were acquired in order to reconstruct a time-variable cycle of 1200. But given the effective rank of five, the reconstruction problem is actually over-determined, though the effective degree of undersampling does not map uniformly across the singular images.

By its nature, the sequence does not probe all degrees of freedom available. This may explain the low rank and also the inability to measure more parameters to high precision. Nevertheless, we were able to measure some important parameters such as $f$ and $T_{1D}^s$. Estimation accuracy is hard to assess because no 'gold standard' measures exist (particularly for ihMT) but our findings are similar to those reported elsewhere. Though $T_1^f$ was also estimated, analysis showed that this parameter may be more biased due to the fixing of other relaxation times in the current implementation. Beyond this proof-of-concept demonstration, future work will consider combining the multiband pulses used here with more normal MRF sequences that also modulate flip angle and include other contrast preparation pulses. Modulating RF frequency offsets may allow us to disentangle the interdependence between $T_{1D}^s$ and $T_2^s$. These sequences could be optimized for precision (rather than maximum ihMT contrast) using established CRLB based methods (28) to boost parameter estimation capability, and a more extensive analysis could be performed to identify and reduce remaining estimation biases.

## 6. CONCLUSIONS

This work presents a cyclic-steady-state sequence using multiband pulses with modulated off-resonance power. As presented, the method is a proof-of-concept that uses the machinery of modern MR fingerprinting on a sequence with constant flip angle, that should not be able to perform any parameter estimation if the normal Bloch Equations model is used. We show that MT-mediated signal changes are actually large and can be used to permit semiquantitative and limited quantitative parameter estimation. Future embodiments of this method will involve more general integration with MRF, using the multiband pulse off-resonance bands as an additional parameter to vary alongside others already used for fingerprinting.

**Data Availability Statement**: The code used to generate the simulation results in this study can be found at: https://github.com/mriphysics/MT-MRF (hash 068cbda was used for the presented results).

# Supporting Information 1: Steady-State Signal Calculation

The time-evolution of magnetization can first be written as Eq. S1 and then reformulated as Eq. S2, where $\mathbf{M} = \begin{bmatrix} M_x^f & M_y^f & M_z^f & M_z^{s1} & M_z^{s2} & M_D^{s2} \end{bmatrix}^T$.

$$\dot{\mathbf{M}} = \mathbf{A}\mathbf{M} + \mathbf{b} \qquad [S1]$$

$$\dot{\widetilde{\mathbf{M}}} = \widetilde{\mathbf{A}}\widetilde{\mathbf{M}} \qquad [S2a]$$

$$\widetilde{\mathbf{M}} = \begin{bmatrix} \mathbf{M} \\ 1 \end{bmatrix} \qquad [S2b]$$

$$\widetilde{\mathbf{A}} = \begin{bmatrix} \mathbf{A} & \mathbf{b} \\ \mathbf{0} & \end{bmatrix} \qquad [S2c]$$

Here, 1 is a scalar value and **0** represents a row vector matching the dimensionality of **A** and **b**. The resultant equation is homogeneous and the solution over time increment TR (during which $\widetilde{\mathbf{A}}$ is constant) is: $\widetilde{\mathbf{M}}(t + \text{TR}) = \widetilde{\mathbf{A}}(t)\widetilde{\mathbf{M}}(t)$. For MT-MRF, we alternate between RF pulses with one, two and three bands. For a given RF pulse, the effect on the magnetization is:

$$\mathbf{R} = \exp(\langle \mathbf{\Omega} \rangle \tau) \qquad [S3]$$

where $\mathbf{\Omega}$ is defined fully by Equation 4 in Malik *et al.* (1) and the angle brackets indicate a time average is taken over the RF pulse of duration τ. Note that here, 'exp' is a matrix exponential. After each pulse, we must apply the operation **S** to account for relaxation and exchange processes:

$$\mathbf{S} = \exp\left(\begin{bmatrix} \mathbf{\Lambda} & \mathbf{C} \\ \mathbf{0} & \end{bmatrix} \text{TR}\right) \qquad [S4]$$

where again, $\mathbf{\Lambda}$ and **C** are defined in Malik *et al.* (1) (Equation 2). Lastly, we can apply the operation $\mathbf{D} = \text{diag}[-1 \ -1 \ 1 \ 1 \ 1 \ 1 \ 1]$ to account for bSSFP phase alternation.

The MT-MRF sequence consists of a repeating cycle of $n_{MB}$ 2B pulses, $n_{1B}$ 1B pulses, $n_{MB}$ 3B pulses, and $n_{1B}$ 1B pulses such that the total cycle duration is $N_{\text{cycle}} = 2(n_{1B} + n_{MB})$. Taking the beginning of the 2B period as the reference point, steady-state magnetization at this point can be calculated by enforcing the periodic boundary condition:

$$\widetilde{\mathbf{M}}(t + N_{\text{cycle}}\text{TR}) = (\mathbf{R_{1B}DS})^{n_{1B}}(\mathbf{R_{3B}DS})^{n_{3B}}(\mathbf{R_{1B}DS})^{n_{1B}}(\mathbf{R_{2B}DS})^{n_{2B}} \widetilde{\mathbf{M}}(t) \qquad [S5]$$

where $\mathbf{R_{1B}}$ corresponds to 1B pulses, etc. This equation can be solved by identifying the eigenvector of matrix $(\mathbf{R_{1B}DS})^{n_{1B}}(\mathbf{R_{3B}DS})^{n_{3B}}(\mathbf{R_{1B}DS})^{n_{1B}}(\mathbf{R_{2B}DS})^{n_{2B}}$ with eigenvalue 1, as described in (1).

# Supporting Information 2: Multiband Pulse Corrections

MT-MRF switches between single-band and multiband pulses with the aim that on-resonance components of these pulses are invariant. When conducting non-phase-encoded experiments, we noticed that significant and unexpected signal fluctuations occurred at the point where pulse type changed and found this was caused by the on-resonance lobes having a slightly different amplitude and phase. A pick-up coil measurement utility already implemented on the scanner was used to characterize these discrepancies for each scan. The origin could not be found but was suspected to be from the RF hardware. Errors were not easily predictable from scan parameters but were found to be consistent as long as scan parameters were unchanged.

Example measurements are plotted in Figure S1 that displays the measured RF pulses in the frequency domain. When zooming in on the on-resonance band, it is clear that there are some discrepancies between the pulses. Complex scaling factors were calculated to make the behavior of each multiband pulse (i.e. 2B and 3B) match that of the single-band (1B) pulse at $\Delta = 0$. The MT-MRF pulse sequence was implemented to allow user-defined values to be entered for these scaling factors. Table S1 shows examples of the required scaling factors.

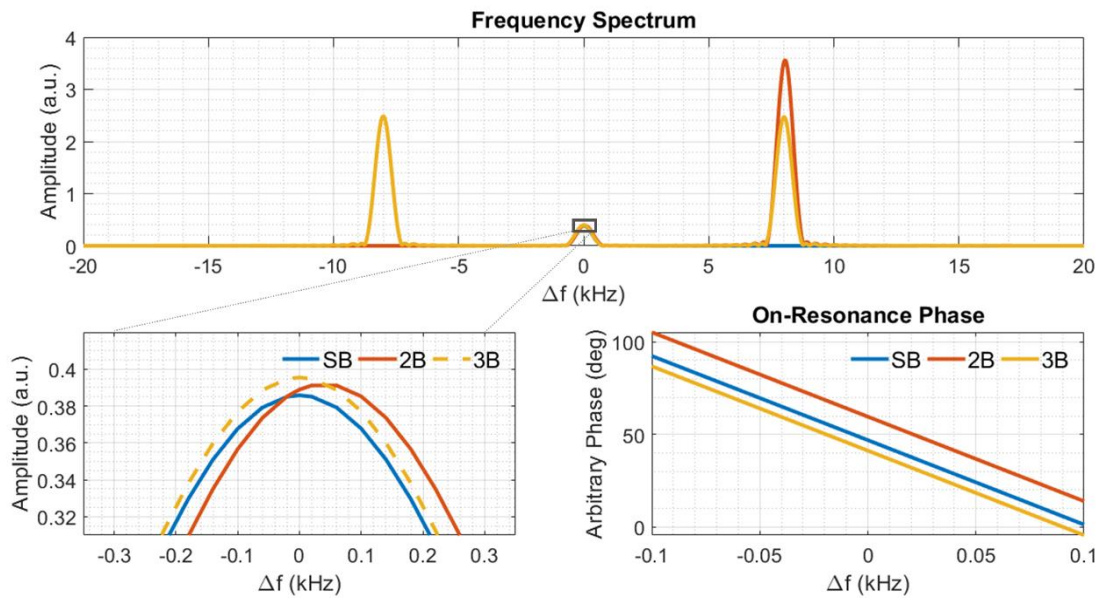

Figure S1: Plots used to derive amplitude and phase correction factors between single-band (SB), 2-band (2B) and 3-band (3B) pulses. Note how the on-resonance lobes and phases of each pulse are slightly offset from one another. These offsets are overcome using the example factors shown in Table S1.

Table S1: Summary of multiband scaling factors used during the phantom and *in vivo* experiments.

| **2B Phase Factor (°)** | **3B Phase Factor (°)** | **2B Amplitude Factor** | **3B Amplitude Factor** |
|---|---|---|---|
| -12.7 | 5.7 | 1.008 | 1.025 |

Non-phase-encoded phantom scans were then repeated to ensure that fluctuations were removed, and this is confirmed in Figure S2 (below).

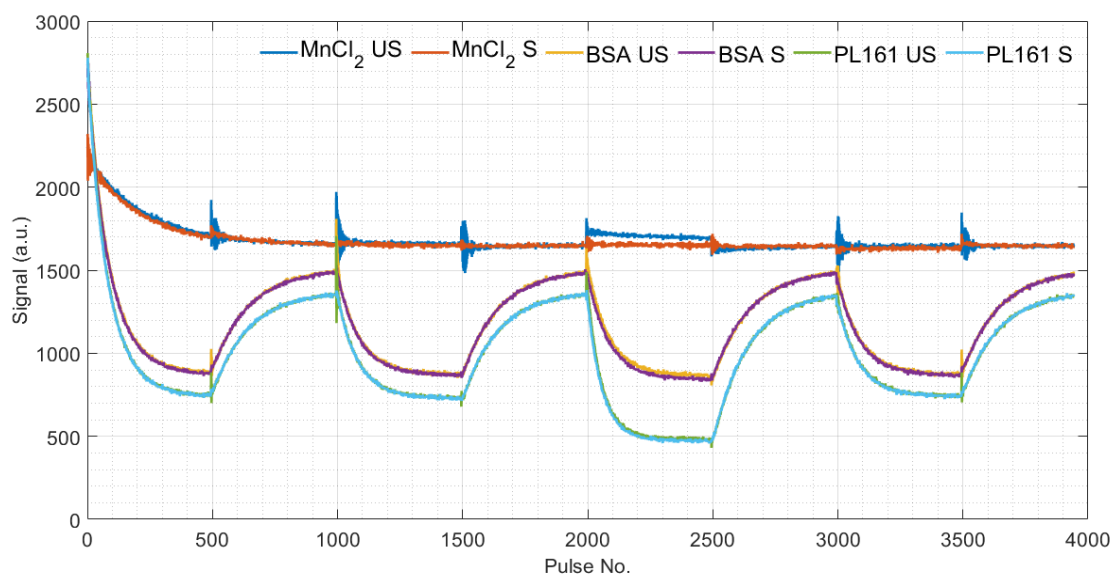

Figure S2: Example 1D signal profiles from each phantom before (unscaled, US) and after (scaled, S) amplitude and phase correction. Large fluctuations at transitions between pulse types are significantly reduced and removed for BSA and PL161. The pulse cycle used to obtain these plots is: {500 $2B^+$, 500 SB, 500 $2B^-$, 500 SB, 500 3B, 500 SB etc.} so does not match the actual acquisition scheme used in the manuscript; it was used for debugging purposes only. $2B^+$ refers to a 2-band pulse with a positive off-resonance lobe and $2B^-$ has an equal but opposite negative off-resonance lobe. Though the latter is not used in our phantom and *in vivo* experiments, it can be used to generate ihMT contrast when combined with a matched $2B^+$ pulse to approximate dual frequency off-resonance.

## Supporting Information 3: Supplementary Phantom Results

To supplement Table 2 in the manuscript, below are single-slice maps resulting from the two separate dictionary fits to MT-MRF phantom data. Note the similarity between MT parameter estimates when either $T_2^f = 84$ms or $T_2^f = 130$ms. Only $T_1^f$ values change slightly due to an expected $T_2^f$ dependence.

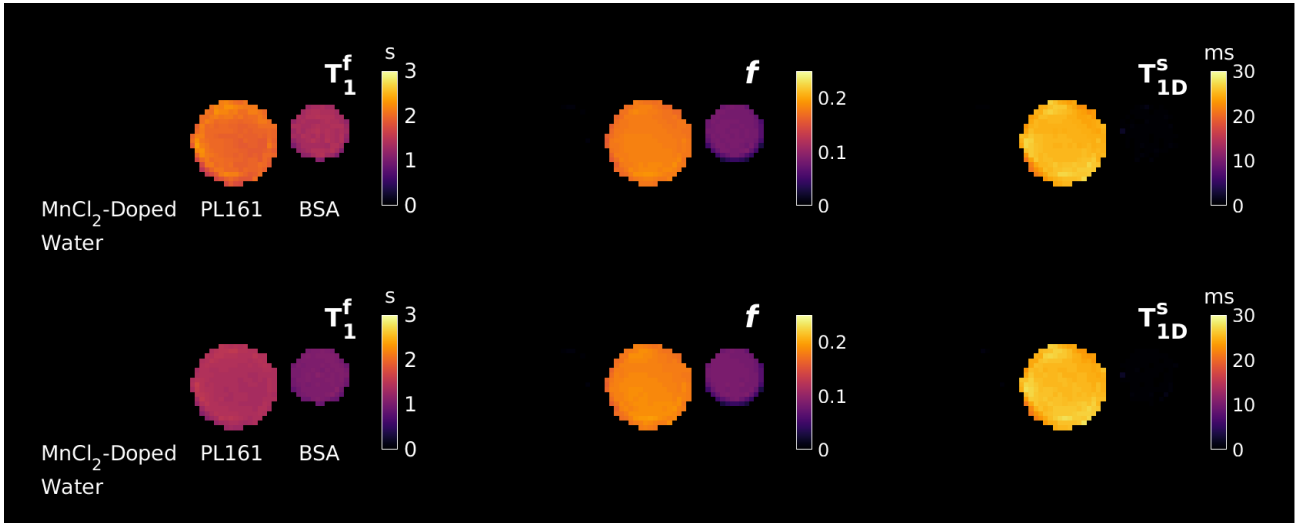

Figure S3: *Top*: MT-MRF parameter maps obtained using a GM-WM average parameter set with $T_2^f = 84$ms (Section 3.2). *Bottom*: Equivalent parameter maps but assuming a more 'PL161-like' fixed value for $T_2^f$ (130ms) only.

## Supporting Information 4: Supplementary *in vivo* Results

To supplement the inter-subject ihMTR comparison shown in Table 1 of the manuscript, below is an equivalent comparison of corresponding MTR values between MT-MRF and ss-ihMT.

Table S2: MTR values obtained from different regions-of-interest for the axial slices of the five subjects in Figure 6.

| MTR (%) | ss-ihMT | | | MT-MRF | | |
|---|---|---|---|---|---|---|
| | **CST** | **Frontal WM** | **Cortical GM** | **CST** | **Frontal WM** | **Cortical GM** |
| **Subject 1** | 47.4 ± 1.8 | 46.6 ± 0.8 | 42.5 ± 1.9 | 36.7 ± 0.7 | 38.4 ± 0.9 | 32.0 ± 2.3 |
| **Subject 2** | 47.4 ± 1.5 | 46.6 ± 0.7 | 44.7 ± 0.6 | 36.1 ± 0.5 | 37.1 ± 0.8 | 33.4 ± 0.9 |
| **Subject 3** | 46.1 ± 1.9 | 45.6 ± 0.9 | 42.3 ± 1.6 | 35.3 ± 0.9 | 37.5 ± 0.9 | 31.8 ± 1.0 |
| **Subject 4** | 47.6 ± 0.8 | 46.8 ± 0.9 | 42.3 ± 0.8 | 36.1 ± 0.4 | 37.0 ± 0.9 | 31.7 ± 1.4 |
| **Subject 5** | 46.6 ± 0.9 | 46.1 ± 1.1 | 42.4 ± 1.7 | 37.1 ± 1.5 | 37.2 ± 1.1 | 31.8 ± 1.7 |

For Subject 2 (male, aged 25), two identical MT-MRF scans were completed approximately eight months apart. Figure S4 presents a central axial slice from each dataset following registration. Table S3 compares measurements in corresponding matched regions-of-interest.

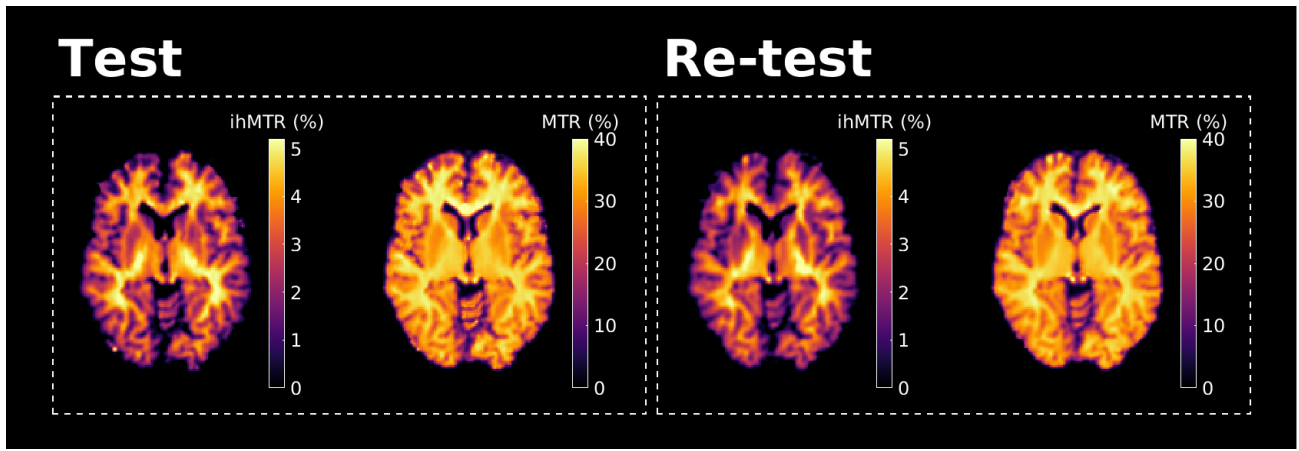

Figure S4: Equivalent central axial slices from the same healthy subject but two separate acquisitions. ihMTR and MTR contrasts from MT-MRF are repeatable according to the region-of-interest measurements shown in Table S3.

Table S3: ihMTR and MTR values from three different brain regions in the axial slices shown in Figure S4.

|  | ihMTR (%) | | | MTR (%) | | |
| --- | --- | --- | --- | --- | --- | --- |
|  | CST | Frontal WM | Cortical GM | CST | Frontal WM | Cortical GM |
| **Test** | 5.02 ± 0.21 | 4.72 ± 0.29 | 3.06 ± 0.25 | 35.4 ± 0.7 | 37.9 ± 0.9 | 31.8 ± 1.5 |
| **Re-test** | 4.98 ± 0.51 | 4.75 ± 0.13 | 2.95 ± 0.27 | 35.8 ± 0.7 | 37.3 ± 1.0 | 31.9 ± 1.1 |

# Supporting Information 5: Additional Dictionary Fitting Results

*Fixing $T_{1Z}^s$ to a Lower Value*

Recently, Wang *et al.* suggested that $T_{1Z}^s$ is much lower than is usually assumed in MT literature. (2) Therefore, we repeat dictionary fits for Subject 1 using a lower, brain-average $T_{1Z}^s$ = 200ms.

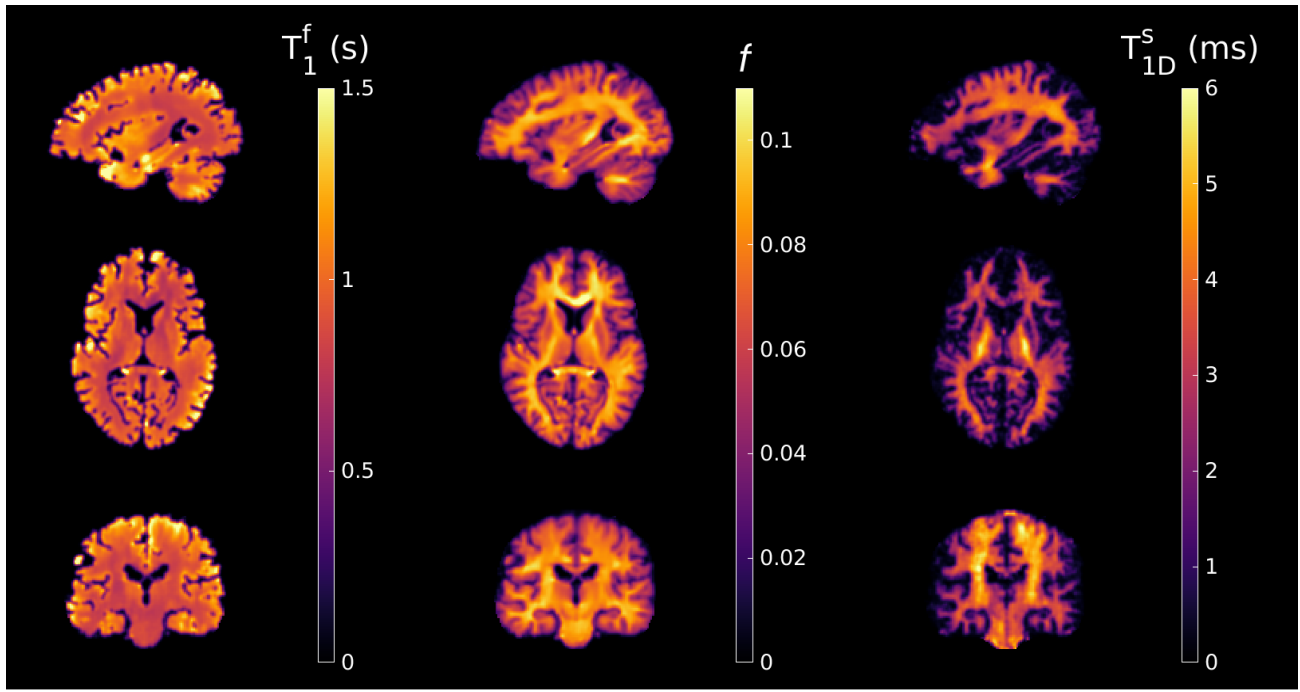

Figure S5a: Dictionary fitting results for Subject 1 assuming lower $T_{1Z}^s$. Compared to Figure 8, $T_1^f$ estimates increase since free and semisolid $T_1$ are coupled to one another, whilst estimates for $f$ and $T_{1D}^s$ are mostly unchanged.

*Fitting for $T_2^f$*

From Figure 3, Combination 4 seems to also give reasonable estimation precision and so dictionary fits are repeated for Subject 1 but for the case where $T_2^f$ is estimated and $T_1^f$ is fixed at 1.36s. (3)

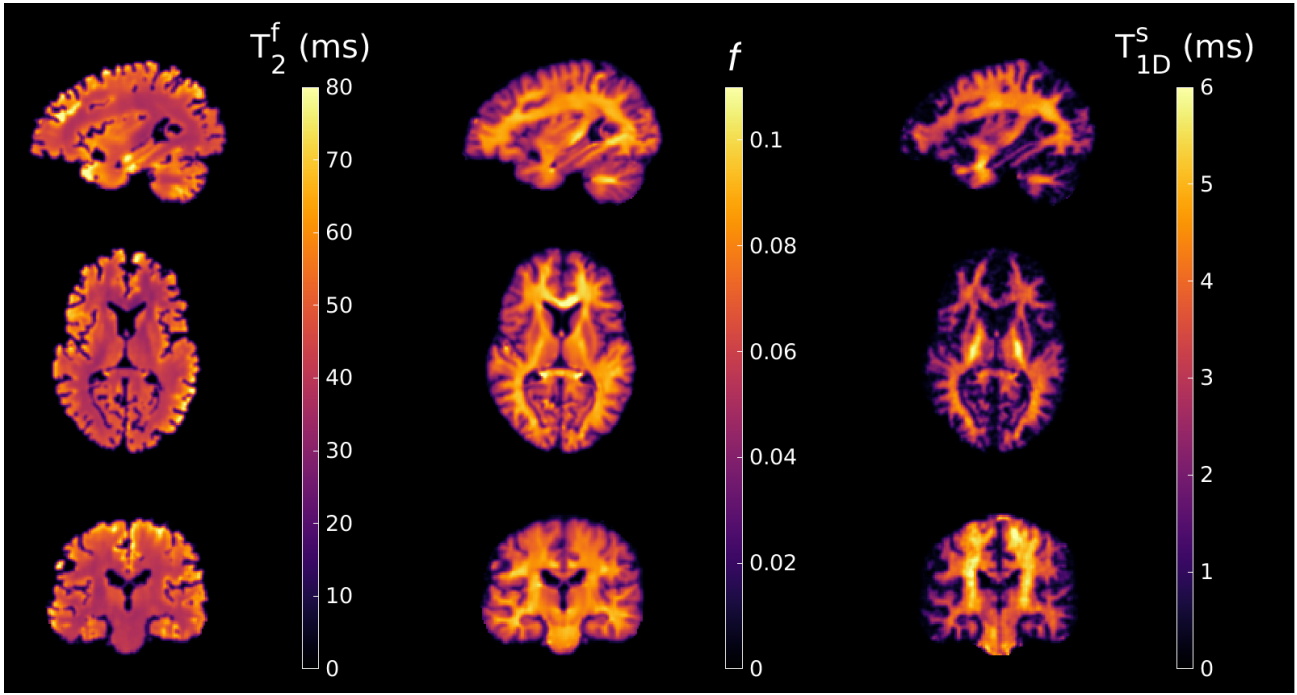

Figure S5b: Dictionary fitting results for Subject 1 assuming fixed $T_1^f$ but estimating $T_2^f$. $T_2^f$ maps show similar contrast to those for $T_1^f$ but once more, estimates for $f$ and $T_{1D}^s$ are very similar to those reported in Figure 8.

Dictionary fitting was also performed by incrementing our previously assumed fixed parameter values by ±10%. Difference maps were then computed between these two to reveal possible interdependence of model parameters and investigate the sources of bias in our *in vivo* dictionary fits.

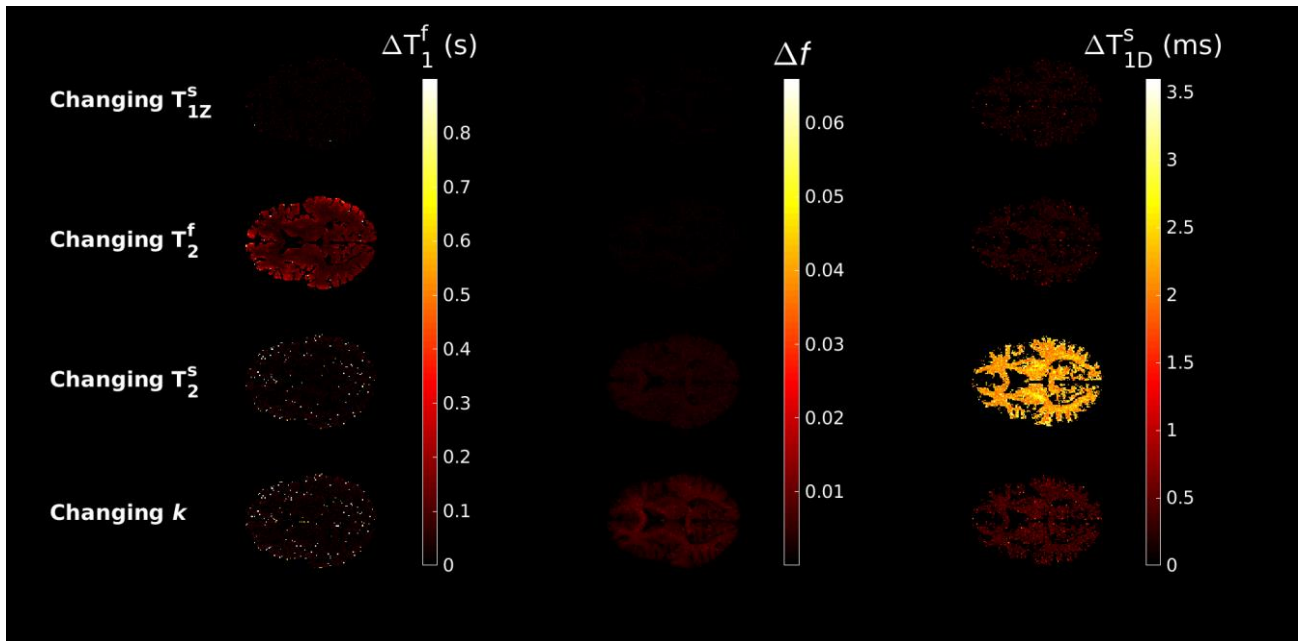

Figure S5c: Difference maps for each estimated parameter given a ±10% change in the fixed parameters stated on the left-hand side. Each estimated parameter seems invariant to small changes in $T_{1Z}^s$; $T_1^f$ is most coupled to $T_2^f$; $T_{1D}^s$ shows considerable dependence on $T_2^s$; and $f$ is slightly influenced by $k$. Colorbars for each estimated parameter change (ΔX, where X is the estimated parameter) are chosen to cover 60% of the ranges used for these quantities in other figures.

## Supporting Information References